\documentclass[conference]{IEEEtran}
\usepackage{cite}
\ifCLASSINFOpdf
  \usepackage[pdftex]{graphicx}
  \graphicspath{{/Users/Sean/Desktop/audio-side-channel/usenix/pics/}}
  \DeclareGraphicsExtensions{.pdf,.jpeg,.png,.eps}
\else
\fi
\usepackage[cmex10]{amsmath}
\usepackage{algorithmic}
\usepackage{array}
\usepackage{mdwmath}
\usepackage{mdwtab}
\usepackage{eqparbox}
\usepackage[font=footnotesize, caption=false]{subfig}
\usepackage{stfloats}
\usepackage[hyphens]{url}
\usepackage{hyperref}
\usepackage[hyphenbreaks]{breakurl}
\usepackage{color,soul}
\usepackage{epsfig}
\usepackage{multirow}
\usepackage{filecontents}
\usepackage{hhline}

\usepackage{filecontents}

\begin{document}
\title{Using Inaudible Audio and Voice Assistants to Transmit Sensitive Data over Telephony}

%\author{\IEEEauthorblockN{Michael Shell}
%\IEEEauthorblockA{Georgia Institute of Technology\\
%someemail@somedomain.com}
%\and
%\IEEEauthorblockN{Homer Simpson}
%\IEEEauthorblockA{Twentieth Century Fox\\
%homer@thesimpsons.com}
%\and
%\IEEEauthorblockN{James Kirk\\ and Montgomery Scott}
%\IEEEauthorblockA{Starfleet Academy\\
%someemail@somedomain.com}}


\author{\IEEEauthorblockN{Zhengxian He\IEEEauthorrefmark{1},
Mohit Narayan Rajput\IEEEauthorrefmark{1},
Mustaque Ahamad\IEEEauthorrefmark{1}}
\IEEEauthorblockA{\IEEEauthorrefmark{1}Georgia Institute of Technology\\
\{zhe72, mrajput7\}@gatech.edu, mustaq@cc.gatech.edu}
}


\IEEEoverridecommandlockouts
%\makeatletter\def\@IEEEpubidpullup{6.5\baselineskip}\makeatother
%\IEEEpubid{\parbox{\columnwidth}{
%    Network and Distributed Systems Security (NDSS) Symposium 2021\\
%   21-24 February 2021, San Diego, CA, USA\\
%    ISBN 1-891562-61-4\\
%    https://dx.doi.org/10.14722/ndss.2021.24xxx\\
%    www.ndss-symposium.org
%}
%\hspace{\columnsep}\makebox[\columnwidth]{}}

\maketitle

\begin{abstract}
%\boldmath
New security and privacy concerns arise due to the growing popularity of voice assistant (VA) deployments in home and enterprise networks. A number of past research results have demonstrated how malicious actors can use hidden commands to get VAs to perform certain operations even when a person may be in their vicinity. However, such work has not explored how compromised computers that are close to VAs can leverage the phone channel to exfiltrate data with the help of VAs. After characterizing the communication channel that is set up by commanding a VA to make a call to a phone number, we demonstrate how malware can encode  data into audio and send it via the phone channel. Such an attack, which can be crafted remotely, at scale and at low cost, can be used to bypass network defenses that may be deployed against leakage of sensitive data. We use Dual-Tone Multi-Frequency (DTMF) tones to encode arbitrary binary data into audio that can be played over computer speakers and sent through a VA mediated phone channel to a remote system. We show that modest amounts of data (e.g., a kilobyte) can be transmitted with high accuracy with a short phone call lasting a few minutes. This can be done while making the audio nearly inaudible for most people by modulating it with a carrier with frequencies that are near the higher end of the human hearing range. Several factors influence the data transfer rate, including the distance between the computer and the VA, the ambient noise that may be present and the frequency of modulating carrier. With the help of a prototype built by us, we experimentally assess the impact of these factors on data transfer rates and transmission accuracy. Our results show that voice assistants in the vicinity of computers can pose new threats to data stored on such computers. These threats are not addressed by traditional host and network defenses that are deployed against data theft. We briefly discuss possible ways in which such threats may be mitigated.
\end{abstract}

\section{Introduction}

Voice offers a natural way for people to interact with devices in their environment which has led to an ever increasing number of computing
devices with embedded microphones and speakers. At the same time, voice assistants (VAs), which facilitate such voice interactions, have become commonplace. For example,
according to a report in early 2019, more than 100 million Amazon Alexa or Alexa enabled devices were sold~\cite{alexa:2019}. Although Amazon Echo and other similar devices (e.g., Google Home) are 
commonly deployed in the home environment, increasingly they are also being used in business settings. Amazon Alexa Business has advocated the use of VAs
for employee work areas and conference rooms. %These devices are connected to the same enterprise networks which connect company computer systems.

Although VAs enable more natural interactions, their connectivity to the network opens them to a variety of new potential attacks. In recent research~\cite{diao2014your, jang2014a11y, Carlini:2016vl, schonherr2018adversarial, Zhang:2017kl, Roy:2018ws, hanspach2013covert}, a number of attacks on VAs have been investigated (see Section~\ref{sec:related}). These attacks primarily focus on issuing malicious
commands to VAs and have also explored how such commands can be made hidden or inaudible so they are not noticed by humans who are close to the 
devices. These attacks provide important insights into how such devices can be exploited by attackers but they do not address two important questions. First, can such attacks
be carried out by a remote actor at scale with low cost? Second, 
%in well defended networks \hl{One reviewer doubted this assumption, as it is hard to realize a perfect defended network in a common setting.}, 
when computers are in close proximity of VAs, can the VA be 
used to exfiltrate sensitive data from these computers? We investigate attacks where VAs are abused as a bridge between the Internet and telephony networks to bypass data leakage protection implemented as a network defense.

As a concrete example in the home setting, consider a laptop or a home office desktop that is close to a VA. Such computers can be infected
with malware with a variety of attack vectors including drive-by-downloads, social engineering and others. Once infected, the malware can choose to locate sensitive data such as credit card numbers or passwords on the  computer and try to exfiltrate it. Although it may not be the case in home environment, we assume that host and network defenses that are employed make it challenging for the malware to connect with the remote host where it wants to drop the data. However, another possibility for the malware is to encode the data into an audio file. The
infected computer can issue a command to a VA to make a phone call and then play the encoded file audio via its speaker. The phone number to which the call is
made can be controlled by the attacker and the call audio could be recorded. From such recorded audio, the data can be decoded. Thus, in this case, data can be exfiltrated 
from a compromised computer even in the presence of effective network defenses. Since malware infections can target a large number of users at very low cost, such attacks can
be crafted at scale and remotely without being in close proximity of VAs.

We leverage prior research in attacks on VAs that demonstrate the feasibility of issuing various commands. Unlike past work, we do not require that the attacker be in the
same physical location as the targeted VA. Also, we utilize the phone channel for data exfiltration after the VA is commanded to make a phone call. Although
past work has explored audio networking and audio covert channels extensively~~\cite{hanspach2013covert, deshotels2014inaudible, madhavapeddy2005audio, nandakumar2013dhwani, kannan2012low, lee2015chirp, guri2017acoustic}  where data is transmitted in audio form between devices via speakers and microphones, we investigate the efficacy of such data transfer over a phone channel that is set up by a VA. Also, similar to hidden commands, we need to make data transmission unnoticeable when a user may be in the same area as the
infected computer or the VA. The challenge thus is to transmit data via audio that is unlikely to be noticed by a human while ensuring accuracy and high transmission rate. Modulation with an ultrasonic  carrier has been proposed to meet this goal~\cite{Zhang:2017kl} but this requires high quality speakers that are not available on computers commonly used in the home and office environments. %\hl{change to about the high frequency modulation.}  
We demonstrate that it is possible to use widely available speakers that come with home and office computers when modulation is done with a carrier frequency close to the higher end of human hearing range which provides a tradeoff between inaudibility and transmission bandwidth.
We explore a concrete setting to explore these questions to quantify the accuracy and bandwidth of data transmission from an infected computer to a remote attacker controlled computer via the Alexa voice assistant. 

Since malware and targeting of computers by it has been investigated extensively, in this paper we focus on the exploitation of VAs by such malware for data exfiltration. More specifically, we explore if a VA can be commanded to make a phone call and if sound, which is inaudible to most people, can be used to encode and  send data over such a call. Since the call may go over a telephony network, only the voiceband (300--3400 Hz) could be used for such data transfer. This motivated our use of Dual-Tone Multi-Frequency (DTMF)~\cite{battista1963signaling} encoding of data by choosing frequencies that avoid overlapping harmonics. Since we modulate the audio with a carrier frequency that is outside of the voiceband,  the telephony channel will act as a low pass filter and drop frequencies higher than 3400 Hz. However, because of VA microphone nonlinearity, encoding DTMF frequencies are reintroduced in the audio sent over the telephony channel. We demonstrate that presence of such frequencies is sufficient to decode the data in the received audio while maintaining the stealthiness of the audio transmission between the computer speaker and the voice assistant microphone.
% so it is nearly  inaudible to humans, we utilize vary high frequency modulation techniques.} After we studied the frequency response of the call path \hl{and the nonlinearity of the microphones on the voice assistant.}, we found this indeed is the case which motivated our  
We explored
the many parameters that can impact the efficacy of transfer of audio encoded data when modulated with a very high frequency carrier and sent over a phone channel via a VA. These include distance between the infected computer and VA, background noise level and stealthiness of the audio based on modulating carrier frequency. 

By developing a system and experimentally evaluating it, we are able to demonstrate that VAs do create a new 
vulnerability for data theft which could be a serious concern in enterprise environments. More specifically, we make the following
contributions in this paper.

\begin{itemize}
\item We demonstrate the feasibility of data exfiltration from infected computers that are in close vicinity of a VA. The data can be sent via the
computer speaker to a VA and then over a phone call. Such attacks can be carried out at scale by a remote malicious actor who is able to commandeer a large number of computers that are in the vicinity of VAs. Thus, unlike much of the past work, we show this can be done without the attacker being in the same physical location as the 
VA.
\item We show that by carefully choosing DTMF and carrier frequencies, we can make the sound that transmits the data inaudible to most people while leveraging voice assistant microphone and phone channel characteristics to naturally recover the encoding frequencies and decode the sent data.
\item We experimentally estimate the bandwidth of the channel that is set up to exfiltrate data. We show that a kilobyte of data can be transmitted with high accuracy by a call lasting less than 5 minutes in a realistic setting where the VA may be several feet away from the computer.
\item We explore the impact key parameters such as stealthiness and distance between VAs and infected machines have on the bandwidth of such exfiltration channels. 
%\hl{We also study the impact of sound masking on such bandwidth.}

\item We discuss the effectiveness of a number of possible defenses against audio encoded data exfiltration attacks. These include liveness detection for voice assistant command audio and monitoring of outgoing audio stream that is fed to computer speakers.
\end{itemize}

\begin{figure}[!t]
\centering
\includegraphics[width=3.1in]{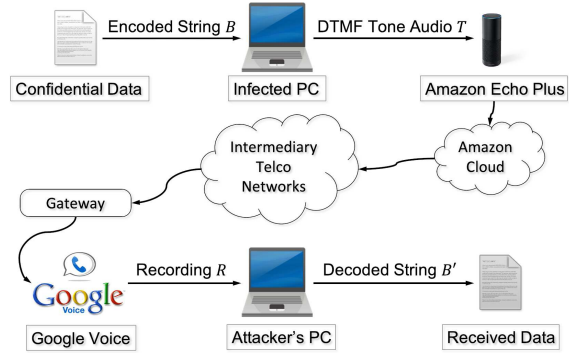}
\caption{The Data Flow of Our Scheme.}
\label{Fig:system}
\end{figure}

\section{Related Work}
\label{sec:related}
%In this section we discuss previous work done to attack voice assistant model. We also discuss some of the defense that is already in place by the manufacturer of the voice assistants to mitigate the threats provided by these methods.
Security and privacy issues related to VAs have been explored by many researchers in the recent past. These efforts have focused on how unauthorized and potentially
malicious commands can be used to target such devices~\cite{diao2014your, jang2014a11y, Carlini:2016vl, schonherr2018adversarial, Carlini:2018wj, Abdullah:2018ho, Yuan:2018um, kumar2018skill, qin2019imperceptible}. Diao et~al.~\cite{diao2014your} demonstrated that an attacker could inject synthetic audio or play commands to control a smartphone. However, the commands could also be noticed by people close to VAs. Carlini et~al.~\cite{Carlini:2016vl} first demonstrated that such commands could be made unnoticeable to humans while VAs would respond to them in the expected manner. They used the term \textit{hidden commands} and a key contribution of this work was to expand the threat model to include humans being in the vicinity of the VAs that were targets of such malicious commands

In another technique that allows a command to be inaudible to humans but recognized by VAs, researchers have used ultrasonic sound carriers to hide the commands from being heard by human~\cite{Zhang:2017kl, Roy:2018ws, hanspach2013covert, deshotels2014inaudible, yan2019feasibility, zhou2019hidden, bispham2019attack, Iijima:2019fh, Yan:2020fa}. Zhang et~al.~\cite{Zhang:2017kl} modulated voice commands on ultrasonic carriers (i.e., $\mbox{frequency} \ge 20\mbox{ kHz}$) to achieve inaudibility. These high frequency encoded commands are sent to VAs. The microphones of VAs can capture the signal at these frequencies, but people cannot hear them. The nonlinearity of the microphone acts as the demodulator of the sound and allows the commands carried on the ultrasonic sound to be decoded and executed. While Zhang et~al.\ demonstrated this attack when the speaker is within 10--20 inch range of the device's microphone, Roy et~al.~\cite{Roy:2018ws} extended the attack range of this ultrasonic sound to 25 ft.\ with a speaker array. Zhou et~al.~\cite{zhou2019hidden} successfully demonstrated how an autonomous car can be controlled with hidden commands. Yan et~al.~\cite{Yan:2020fa} demonstrated how a wave generated by an ultrasonic transducer can propagate in a solid material (e.g.\ a table.) to reach the VA. These attacks, which rely on ultrasonic sounds, all require speakers of high quality or additional equipment in close proximity of the target. In our work, we do not use ultrasonic sound modulation because it cannot be effectively transmitted via speakers that are commonly available in computers. We also require that the modulated sound can be transmitted over an arbitrary distance to an endpoint controlled by an attacker. We exploit voice assistant microphone nonlinearity induced demodulation to achieve this via the voiceband of a telephony channel. Also, previous work is only focused on issuing malicious commands to VAs, not on using them for data exfiltration over long distances.

Malicious commands directed at VAs can also be hidden in other sounds that may not raise an alarm when heard by humans. Sch{\"o}nherr et~al.~\cite{schonherr2018adversarial} proposed a method to hide malicious command in different sounds like songs and  sounds of chirping birds to take control of VAs.  Such
attacks could be effective because of \textit{psychoacoustic hiding}, which is based on the  fact that when human brain is busy processing a loud sound of a certain frequency, it is likely to ignore other quieter sounds at adjacent frequencies for a few milliseconds. Such adjacent frequencies can be used to hide voice assistant commands in high-frequency audio such as chirping of birds. 

Adversarial machine learning methods have also been explored for hiding malicious audio commands. Carlini et~al.\ used audio adversarial examples to compromise voice recognition systems in~\cite{Carlini:2018wj}. Such work primarily relied  on generating audio files which are then fed directly to a voice transcription system, not transmitting the audio through air to a VA. Abdullah et~al.~\cite{Abdullah:2018ho} and Yuan et~al.~\cite{Yuan:2018um} presented more practical adversarial attack techniques that could be used in a more realistic setting. These works do demonstrate the feasibility of hidden commands against VAs. However, they do not investigate how such commands can be used by attackers to exploit compromised computers.

Covert or side channels based on audio signals have been explored extensively in the past.
%For example, sounds produced by a device can be used as a covert channel in~\cite{guri2017acoustic, hanspach2013covert}.
Guri et~al.~\cite{guri2017acoustic} utilized the noise emitted from the CPU and chassis fans to transmit data bits from computers. Hanspatch et~al.~\cite{hanspach2013covert} presented a covert acoustical mesh network with ultrasonic sounds. Deshotels ~\cite{deshotels2014inaudible} expanded this method to mobile devices. All of these techniques only work at short ranges because either the device sound must be captured directly or microphones must capture the transmitted ultrasonic signal. Our goal is not to explore new audio covert channels but to investigate the threat posed by such channels when VAs are in the vicinity of computers. Furthermore, we address the problem of exfiltration of data via audio to remote computers.

Data transfer with acoustic signals has been explored in the past~\cite{kannan2012low, lee2015chirp, madhavapeddy2005audio, nandakumar2013dhwani}. Madhavapeddy et~al.~\cite{madhavapeddy2005audio} presented the use of Dual-Tone Multi-Frequency (DTMF) signaling for such transfer to send short messages between devices over the phone channel. A bit rate of 20 bit/second with DTMF across a 100-inch distance was demonstrated. Nandakumar et~al.~\cite{nandakumar2013dhwani} realized an acoustics-based near field communication (NFC) system on mobile phones. It used orthogonal frequency division multiplexing (OFDM) with subcarriers selected in range 0--22 kHz to send data. Our proposed system uses DTMF signaling to encode data but we explore its transmission over a telephony channel to attacker controlled remote systems via a VA.

In this paper, we assume that data is exfiltrated from a compromised computer that is close to a VA. The computer is compromised by infecting it with some malware. There is considerable research in the area of malware infection, detection and remediation (see survey papers such as~\cite{egele2012survey,cert2005malware}). One or more of the infection
vectors can be used by malicious actors to infect machines that store sensitive data. We do not discuss details of how such infections are accomplished but focus on how the malware can utilize the speaker on the infected machine and a VA to exfiltrate sensitive data.

%https://www.us-cert.gov/sites/default/files/documents/NCCIC_ICS-CERT_AAL_Malware_Trends_Paper_S508C.pdf
%\hl{Malware infections}

%\hl{The normal internet was under the protection of Intrusion Detection System(IDS) and Intrusion Prevention System(IDS)}

\section{Threat Model}
We explore attacks in which an attacker's goal is to locate computers in home or office settings that store sensitive data and then exfiltrate such data via nearby VAs. To launch such an attack at scale, the attacker must be able to infect large number of computers and take control of them via malware that can be used to issue commands to VAs. Malware infection of victim computers can be accomplished using a number of techniques, including compromised webpages,  spam email clicks, social engineering or by exploiting software vulnerabilities. Once a computer is under control of the malware, the malware must  issue a series of audio commands via available device speakers to activate a nearby VA. Once this is done,  it plays audio via computer speakers that encodes the data and the data then is transmitted to an attacker controlled system. 

Although we assume that the attacker is able to compromise computers via tricking  users, we do not require that the attacker is able to overcome host or network defenses that are deployed to protect transfer of sensitive data to unauthorized hosts. Instead, by using computer speakers and VAs, an alternate path is set up to exfiltrate data. Since the data is encoded as audio, one question is if a human, who is in the vicinity of the infected computer or VA, will hear such audio and be alarmed. Although past research has demonstrated that commands to VAs can be hidden or be made inaudible (this is possible through adversarial example attacks~\cite{Yuan:2018um,Carlini:2016vl,Carlini:2018wj} or audio processing~\cite{Abdullah:2018ho,Roy:2018ws,Zhang:2017kl}), it is not known if this can be done with audio that encodes arbitrary data which is played over speakers that are available on common computers. Furthermore, we need to encode data in audio that can be transmitted with high accuracy over a voice phone call connection. 

%\begin{itemize}
%\item \textbf{User not present in the vicinity of the infected computer or voice assistant:} The infected machine can encode the sensitive data in audio using any available scheme and can play the audio so it can be heard by the voice assistant. The audio can be played at low or moderate volume and there can be a certain level of background noise. In a business setting, this could be the case outside of normal business hours. Also, the malware can use the computer microphone to listen to human sounds and in their absence of such sounds after a period, can assume that no human is present (e.g., residents of a home are away at work during work hours).
%\item 
%\textbf{User present in the vicinity of device:}
A person who hears sound coming from a computer will become suspicious when the sound is due to audio encoded exfiltration of data.  Ideally, we can encode the data in audio frequencies that are inaudible to humans using modulation with an ultrasonic carrier but as discussed earlier, this is not possible with ordinary computer speakers. Instead, we consider a threat model that seeks to achieve near inaudibility based on the following observation. Most adults cannot hear high frequencies even when they are close to the sound source. For example, an average adult can only hear frequencies below 16 kHz due to natural hearing loss with age. As shown in audiograms that capture normal human hearing, the highest frequency sound that can be noticed is even lower (e.g., less than 10 kHz) when the sound volume is at the level of a normal conversation or lower~\cite{Suzuki2003PreciseAF}. We explore if data can be exfiltrated stealthily when  the audio encoding the data is modulated with a very high frequency carrier (over 15 kHz). In fact, as expected, we did find a tradeoff between inaudibility and the accuracy with which the audio can be transmitted via computer speakers. We can increase the carrier frequency to 20 kHz to make it inaudible to nearly everyone at lower transmission accuracy or we can choose a frequency that most people would not hear to increase accuracy. 
We report results of experiments that explore this tradeoff and measure the bandwidth of the audio data transfer channel with various high frequency modulation carriers.This helps us understand data transfer rates that could be achieved with various levels of inaudibility for humans who may be close to the computer.

\section{Data Transmission Channel Characterization}
\label{Sec:phone}
Our goal is to explore if the audio channel that can be setup between a computer and a nearby VA can be leveraged for transmitting data to an attacker controlled remote computer. We use the phone call functionality provided by VAs like Amazon Echo to transmit data to such a remote host. To be able to do this, we need to better understand the characteristics of this transmission channel between the source and target computers. As shown in Figure~\ref{Fig:system}, data is first encoded in an audio file at a compromised source host and this audio is modulated with a high frequency sound carrier to make it less noticeable to a user who may be close to the source computer or the VA. This modulated audio is then transmitted via the computer speaker and received by VA's microphones. Environmental noise could be mixed with the computer transmitted audio. Once the VA captures the audio, a phone call is used to transmit the data to the remote host. Such a call relies on a Voice-Over-IP (VoIP) channel to the cloud and then a gateway that sends the voice over a telecommunications network to the end system where the voice is recorded (a Google Voice phone number is used in our case). 

The phone channel behavior depends on codecs and sampling rates, protocols used for transmitting the audio packets, and other factors such as noise cancellation and packet loss concealment. Although information about audio codecs and transport protocols is available, characterization of the end-to-end transmission channel that starts with a computer speaker and ends with the called phone number is challenging because we have no information about telecommunication carriers and gateways that are involved in supporting a call. Due to this, we take a blackbox empirical analysis approach to study this channel to understand its characteristics that are relevant to transmission of data with audio. In particular, we measure the frequency response of the channel, its transport protocol behavior, and VA features that handle voice capture depending on energy level of the audio source and microphone response to modulation carrier frequencies.

As mentioned earlier, in our experiments, we chose  Amazon Echo Plus as the VA because it can be made to dial a phone number directly by saying, \textit{Alexa, call (123)-456-789}. Thus, it can reach a phone number from any personal Echo. We place the Echo Plus close to a MacBook Pro, playing audio through its speakers. The other endpoint of the phone call channel is a Google Voice number. We record calls coming to the Google Voice number. The audio recordings are downloaded and data is reconstructed from these recordings. Unlike some of the earlier attacks that have been reported on VAs, in our setup a remote attacker can easily get a Google Voice number and download call recordings coming to it. Thus, data can be exfiltrated over the audio channel enabled by the VA without being in physical proximity of the targeted computer and VA.

\subsection{Phone Call Channel Frequency Response}
\label{subsec:frequency}

\begin{figure}[!tbp]
\centering
\includegraphics[width=3in]{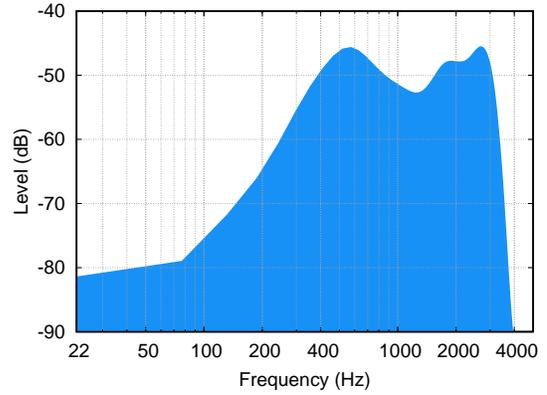}
\caption{Sound level distribution of received audio at various frequencies. To describe signal strength in this paper, we use dB Full Scale
(dBFS), which is based on the ratio of the digital signal level to the maximum possible signal
level~\cite{mason2011terminology}. The values on this scale are different from the air pressure based dB scale that is used to describe sound loudness.}
\label{Fig:freq_range}
\end{figure}

Our data transmission relies on a phone call channel, which is typically used to transmit human voice over the frequency range of 300 to 3400 Hz~\cite{Reaves:2016tq}. To measure if our transmission channel has a similar frequency range, we play a frequency scan audio and measure the sound level of received audio. In Figure~\ref{Fig:freq_range}, the received sound level goes above -60 dB after about 260 Hz, and drops below -60 dB at 3400 Hz, which provides good quality for normal phone call that carries the audio of people talking. While the phone call channel provides limited sound level at lower frequency, it drops sharply after 3000 Hz, and becomes nearly undetectable after 4000 Hz.  Thus, for transmitting data as audio, we must operate within the voiceband (300--3400 Hz).

In theory, the bit rate for data transfer over the phone channel can reach 33600 bit per second~\cite{Reddy:2001kh}, but only through a well designed modem. However, we rely on devices that support normal phone calls. The audio is played out by the speakers on a laptop and captured by microphones on Echo Plus over the air. This audio is then transmitted via a phone channel after encoding it with voice codecs. Also, the quality of the audio transmitted by the computer speakers could be degraded with environmental noise which can further reduce the bandwidth. Thus, the bit rate that our phone call channel can support for data transfer is significantly lower. Although this would make bulk data transfer challenging, passwords, keys, bank account or social security numbers and other limited size data can be transmitted over such a channel.

\subsection{Packet Loss}
\begin{figure}[htbp]
	\centering
	%\subfloat{\includegraphics[width=3in]{pics/packet_loss.eps}}
	%\qquad % here you can insert horizontal or vertical space
	%\subfloat{
	\includegraphics[width=3in]{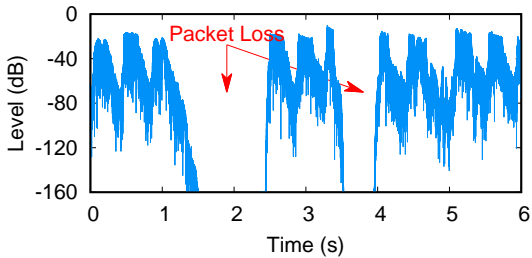}
	%}
	\caption{Packet Loss Shown as Energy Drop.} % caption for whole figure
	\label{Fig:packet_loss}
\end{figure}

VoIP systems often choose the user datagram protocol (UDP) because it provides better call experience and loss of occasional packets can be handled without negative impact on call quality (e.g., packet concealment techniques have been developed to deal with packet loss). However, if we encode data in audio, packet loss could impact accuracy of data transmission. We ran experiments to see if we observe packet losses in the data transmission channel used by us. Figure~\ref{Fig:packet_loss} shows 6 seconds of audio transmitted through our phone call channel. We do observe random packet drops which correspond to the energy level drops in the audio signal. Packet loss may be due to a weak network connection, congestion etc. 

In case of packet loss, the VoIP system may employ techniques like packet loss concealment and such modified audio packets will introduce errors in the decoded data. Since we only check audio after it is recorded at the receiving end, the presence of multiple packet losses could indicate that the data sent in the audio may not be recoverable. Although correct and complete data exfiltration is not achieved in this case, the reconstructed data may provide partial information about the transmitted secret. We will address how data transfer accuracy can be improved in the presence of noise and packet loss in Section~\ref{Sec:method}.

\subsection{Voice Assistant Audio Processing Features}
\label{subsec:preprocess}
\begin{figure}[!t]
\centering
\includegraphics[width=2.7in]{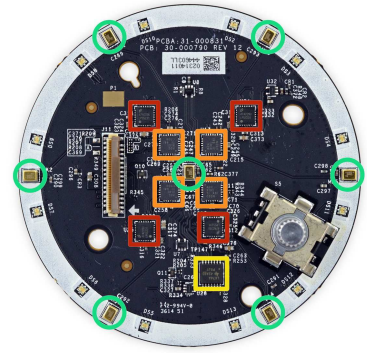}
\caption{Microphones on Amazon Echo Plus in Green Circles. There are six \textit{S1053 0090 V6 Microphones} on the  perimeter and one in the center of the board~\cite{Teardown}.}
\label{Fig:microphones}
\end{figure}

VAs are built for taking people's commands and responding to them. A good quality recording is essential for correct understanding of commands and also for providing a high quality voice call to the other end of phone call channel. Unlike usual cellphones which are only equipped with microphones on the top and bottom, a VA can utilize multiple microphones as a microphone array. In Figure~\ref{Fig:microphones}, there are seven microphones in the top board of Amazon Echo Plus, six on the perimeter and one in the center. A microphone array can help the VA determine the direction of incoming voice based on the difference in the arrival times at each microphone. When audio signals come from many directions, Echo Plus has its own algorithm to distinguish human voice from background noise and then focus in the direction of human voice, resulting in cancellation of noise. In our experiments, we can observe two interesting adjustments Echo Plus applies to its audio recordings.

\begin{figure}[htbp]
	\centering   
	\subfloat[The automatic adjustment of sound level.\label{Fig:adjust}]{\includegraphics[width=2.7in]{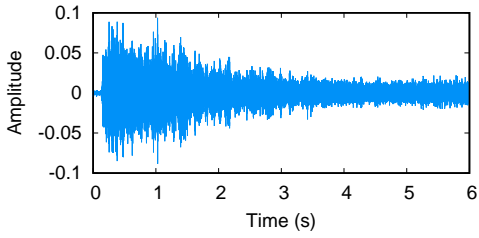}}
	\qquad % here you can insert horizontal or vertical space
	\subfloat[Play same message audio without noise (Before $t = 25$s) and with noise (After $t = 25$s). \label{Fig:attention}]{\includegraphics[width=2.7in]{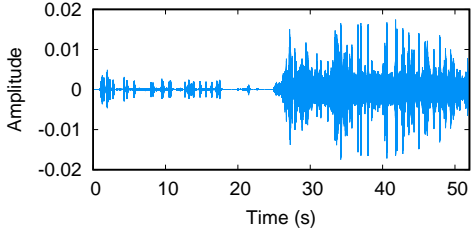}}
	\caption{The Built-in Audio Processing of Amazon Echo Plus.}
\end{figure}

\begin{itemize}
\item Amazon Echo Plus tends to adjust the volume to a moderate level as time goes. From Figure~\ref{Fig:adjust}, while white noise sound level played from laptop is stable, the maximum sound level of the recording is lowered automatically by the device. This may be an adjustment scheme in case of a sudden burst of sound.

\item The sound must reach a certain level to get the \textit{attention} of Amazon Echo. We explore this by conducting an experiment that starts by transmitting audio with no noise at first and then same audio is played with noise. This is repeated to see the activation of the microphones. In Figure~\ref{Fig:attention}, we first show the recorded audio with no noise followed by including the noise. At first, we play our encoded audio from 100 inches away while there is no noise playing. The sound level is too low to catch the \textit{attention} of Amazon Echo Plus. When we play a loud noise with the same encoded audio, the combined sound energy level reaches the threshold and the microphones are activated. We decode the recorded audios which encode the same message. In the case when there is no noise, we could not reconstruct the message. In contrast, the audio with the noise has most of information in it. Although somewhat surprising, this observation is consistent with past research results that the VA is able to understand degraded quality voice commands that are not intelligible for humans.

%\item Amazon Echo Plus will also pay more attention to the side where the sound level is louder.

\end{itemize}

For call audio recorded at the receiving end, we also found that Amazon Echo Plus does not filter out noise from the transmitted audio which was the case when we directly recorded the audio on a cellphone.  %Moreover, sometimes we may need to increase the masking noise level to get the attention of the device.

Based on our characterization of the data transmission channel and the VA from the above experiments, we explore an approach that can help us demonstrate that data can be exfiltrated from a compromised computer via a phone call initiated by an infected computer. Our goal is to achieve high accuracy while making the sound as unnoticeable as possible. We use modulation with a very high frequency carrier to make the sound inaudible and rely on VA microphone nonlinearity for automatic demodulation. We discuss how this can be done next.

\subsection{Voice Assistant Microphone Response to Very High Frequency Sound}\label{subsec:nonlinear}
Very High Frequency Sound (VHFS) is the sound in frequency range 13.5--20 kHz~\cite{doi:10.1121/1.5063818}. We know that sound beyond 20 kHz, referred to as ultrasound, is inaudible to humans. In~\cite{Zhang:2017kl}, Zhang et~al.\ took advantage of ultrasound modulation and the nonlinearity of microphones to send inaudible commands to VAs. They used high quality speakers to transmit ultrasound frequencies. Since our threat model assumes a remote attacker, we only assume speakers that are available in PCs or laptops. While these normal speakers cannot produce ultrasound at a significant energy level, they can generate loud VHFS. If we can modulate the tones with carrier frequency beyond 15 kHz, which is difficult to hear for most adults~\cite{hearingloss}, the modulated sound will likely not raise an alert when people are nearby the computer. Our experiments revealed this to be the case when a person is not right next to the computer speakers.

The VHFS is captured by the voice assistant microphones. The inherent nonlinearity of these microphones results in demodulation which can be modeled as follows\cite{Zhang:2017kl}:
\begin{equation}
s_{out}(t) = As_{in}(t) + Bs_{in}^{2}(t), \label{eq:nonlinearity}
\end{equation}
where $s_{in}(t)$ is the input signal, $s_{out}(t)$ is the output signal, and $A$ and $B$ are gains for corresponding terms.

In our scheme, a two-frequency tone represents a 6-bit binary string (see Section~\ref{Sec:method}).We can formulate the original signal as:
\begin{equation} 
m(t) = \sin(2\pi f_1t) + \sin (2\pi f_2 t), \label{eq:m}
\end{equation}
where $f_1$ and $f_2$ are the two frequencies. Let the carrier frequency to be $f_c >>$ 3400 Hz. We modulate $m(t)$ on $f_c$. Amplitude modulation gives us:
\begin{equation}
y(t) =[1 + \sin(2\pi f_1t) + \sin(2\pi f_2t)]\sin(2\pi f_ct). \label{eq:y}
\end{equation}
By expanding (\ref{eq:y}) and using trigonometric identities, we get five different frequency components: $f_c$, $(f_c - f_1)$, $(f_c - f_2)$, $(f_c + f_1)$ and $(f_c + f_2)$. To keep the signal wave beyond frequency $f_c$ for inaudibility, we filter out $(f_c - f_1)$ and $(f_c - f_2)$. The modulated signal is then sent via the speaker.
\begin{equation}
%\begin{split}
y_s(t)=\sin(2\pi f_c t) - \frac{1}{2}\cos[2\pi(f_c + f_1)t] - \frac{1}{2} \cos[2\pi (f_c + f_2)t].\label{eq:ys}
%\end{split}
\end{equation}
After recording by the microphones, which means applying (\ref{eq:nonlinearity}) to (\ref{eq:ys}), we will get $f_1$, $(f_1 - f_2)$ and $f_2$, but also other unwanted frequencies such as $2f_c$, $2(f_c + f_1)$, $(2f_c + f_1)$, etc. All these high frequencies are automatically filtered out while the recording is sent through the phone call channel with a frequency upper bound of 3400 Hz. We only have $f_1$, $(f_1 - f_2)$ and $f_2$ components remaining, with coefficients $1$, $1/2$ and $1$, respectively. 

\begin{figure}[!t]
\centering
\includegraphics[width=2.7in]{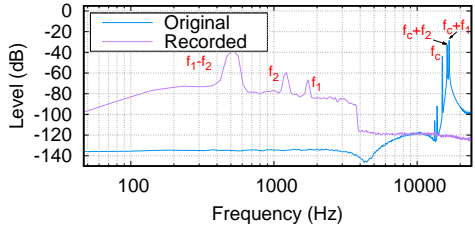}
\caption{Evaluation of the nonlinearity effect. Frequency domain plots for the original audio and recorded audio. In the recoding, $f_1$ and $f_2$ were demodulated out while frequencies beyond about 4000 Hz was filtered out.}
\label{Fig:15k_org}
\end{figure}

In Figure~\ref{Fig:15k_org}, we use a carrier frequency $f_c = 15$ kHz. The appearance of $f_1$, $(f_1 - f_2)$ and $f_2$ shows that the microphones on Amazon Echo Plus do have nonlinearity. We exploit this for decoding data from the audio that is recorded at the receiving end of the phone channel.

\section{Methodology}
\label{Sec:method}
Based on our characterization of the data transmission channel, we explore how an attacker who has control of a computer can exfiltrate data stored on it to a remote system under its control. Since data is encoded in audio, the attacker wants to achieve high accuracy even in the presence of some noise and wants the transfer to complete as quickly as possible. Although several techniques exist for encoding digital data into audio, we focus on \textit{Dual-Tone Multi-Frequency Signaling (DTMF)} because the audio must be transmitted over a phone channel. Also, based on our observations, packet losses and other noise could lead to errors in the received audio. To increase data transmission accuracy, we need to use encoding that could help correct certain class of errors. Finally, to make the transfer stealthy, we modulate the audio that encodes data using a very high frequency carrier so the audio transmission from a computer speaker to the VA is less likely to be noticed by humans. We utilize previously investigated techniques to address data encoding into audio, error handling and high-frequency modulation to demonstrate the feasibility of a stealthy data exfiltration channel to a remote host via a VA. First, we provide a brief overview of these in this section.

\subsection{Preliminaries}
We first introduce DTMF, which we use for transmitting voice encoded data through a phone call channel, and explain how we choose the DTMF frequency groups in our method to increase data transfer rate. We then discuss modulation with a very high frequency carrier in order to make the sound unnoticeable to most humans. 

\subsubsection{Dual-Tone Multi-Frequency Signaling}
DTMF was introduced in~\cite{battista1963signaling} over 50 years ago and is also called ``Touch Tone''. DTMF is specially designed for transmitting digit signals  through a voice call. Since it is difficult to protect single-frequency tones from perturbation of speech and background noise, DTMF uses a set of 16 dual-frequency tones, representing 10 decimal digits and other buttons on a phone. The 16 tones are formed by a so-called ``four-by-four'' code: each tone is made up of one high frequency from \textbf{high~group} and one low frequency from \textbf{low~group} as shown in Table~\ref{Table:code}. Since speech often involves lower harmonics, in order to distinguish DTMF tone frequencies  from speech, Battista et al.~\cite{battista1963signaling} carefully chose them to avoid frequencies in one group harmonically relating to those in the other group. Also, frequencies used in DTMF totally fall within the voiceband range, 300 Hz to 3400 Hz, as we discussed in Section~\ref{subsec:frequency}, to confine them to the transmission band of the telephony channel. DTMF is still widely used for customer services, remote guest door controlling, etc. It is natural and reliable to leverage DTMF for data transmission in our setting that includes a VA being able to make a call.

\begin{table}[!htb]
\centering
\caption{High Group and Low Group of Tone Frequencies.}
\label{Table:code}
\resizebox{.3\textwidth}{!}{%
\begin{tabular}[t]{c|c}
High Group (Hz) & Low Group (Hz) \\ \hhline{=|=}
1209            & 697            \\
1336            & 770            \\
1477            & 852            \\
1633            & 941            \\ 
\end{tabular}%
}
\end{table}

To recover the frequencies used in a tone on receiver side, we make use of the \textit{Goertzel algorithm}. It is an efficient algorithm that can calculate the signal strength at chosen frequencies, unlike fast Fourier transform (FFT), which calculates all frequency components~\cite{oppenheim2014discrete}. 
%Since DTMF only has 8 frequencies to detect, Goertzel algorithm saves a lot of computation time.

\subsubsection{Choosing DTMF Frequencies to Increase Data Transfer Rate}

%\hl{Use the full band. Avoid overlapping.} 
Commonly used DTMF only makes use of 4 low frequencies and 4 high frequencies, as shown in Table~\ref{Table:code},
to form 16 tones. Adjacent frequencies in each group have a ratio of $21/19$. Although these frequency groups are sufficient to transmit 16 symbols on a phone, more tones are needed for our data transmission.  Another problem we found is that background noise includes frequencies that are close to the low frequencies shown in Table~\ref{Table:code}. We choose different DTMF frequencies to address both of these problems.

To increase the number of bits that can be sent in a single tone, we use two expanded frequency groups with 8 frequencies in each group to form 64 tones. A straightforward way to form these groups is to expand the original frequency groups in Table~\ref{Table:code}. However, since the high-frequency modulation we plan to use produces an additional difference frequency $(f_1-f_2)$, a simple expansion while maintaining the desired ratio between consecutive frequencies will produce a lot of collisions between the DTMF and difference frequencies. A collision occurs when two frequencies are very close and the energy level of one could exceed the other one due to noise and leakage due to nonlinearity. This will result in the Goertzel algorithm returning an incorrect tone frequency since it chooses the two frequencies in a tone that have the highest energy levels in each group. For example, if we choose $f_1 = f_A$ from Group A, and $f_2 = f_{B}$ from Group B, the resulting difference frequency $(f_1 - f_2)$ will still be in the voiceband range and could collide with one of the other DTMF frequencies. 

We reduce the likelihood of collisions between tone and difference frequencies by choosing tone frequencies $f_1$ and $f_2$ as follows. We set $f_1 = f_A + f_{B}$ and $f_2 = f_A$, which results in audio in a tone with frequencies  $(f_A + f_{B})$ and $f_A$. This tone audio is modulated with a carrier and transmitted via a speaker. The voice assistant microphone will reintroduce frequencies $(f_A + f_{B})$, $ f_A$ and difference frequency  $f_{B}$ due to its nonlinearity prior to transmission of this audio over the phone channel. A positive side effect of this is that collisions due to the additional higher frequency $(f_A + f_{B})$ in the recording at the destination will be minimized because in most cases, it will be outside of the voiceband and get filtered out by the phone channel which acts as a low pass filter. Based on these insights, we decided to use two groups of frequencies that satisfy the following constraints:

\begin{enumerate} 
\item All frequencies are within 300 Hz to 3400 Hz range.
\item In each group, two adjacent frequencies have a ratio of 21/19.
\item Any two frequencies chosen from the two groups, $f_A$ from one group and $f_{B}$ from the other group, are separated by a threshold. Also, their sum, $(f_A+f_{B})$, is also separated by that threshold from any chosen frequencies. 
\end{enumerate}

The smallest gap between any pair of frequencies  in the original DTMF groups is 73 Hz. However, setting the threshold to 73 Hz, 72 Hz and 71 Hz gives no results that satisfy the above constraints. When we set the threshold to 70 Hz,  we obtain the frequencies shown in Table~\ref{Table:ext_code} that satisfy these constraints. These 16 frequencies form the two groups that allow us to produce $8\times8 = 64$ tones, and each tone can thus encode 6 bits.

 %First, we were empirically able to establish that lower frequencies often appear in noise that is common in settings like office and home. 

%We can extend the frequency set by adding four lower frequencies, $\{468, 517, 571, 631\}$(Hz) below the lowest frequency 697 Hz of the low group of $4\times4$ DTMF. Alternatively, we can add the same number of frequencies, $\{1040, 1149, 1270, 1404\}$(Hz) above the highest frequency 941 Hz of the low group. Both of these keep low group within voiceband. The same can be done for high group. In the absence of background noise, both choices should lead to similar performance for data transmission. However, the frequencies must be chosen to avoid overlap with background noise and from other chosen frequencies.

\begin{figure}[htbp]
	\centering
	\includegraphics[width=3in]{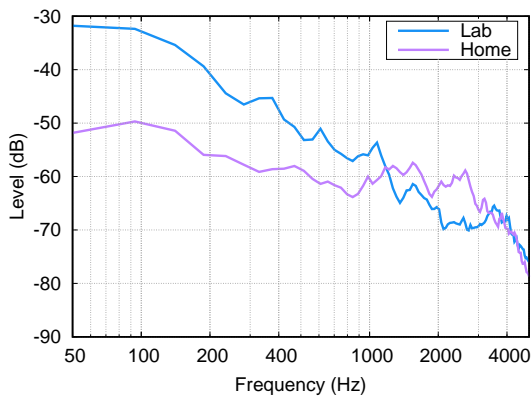}
	\caption{Noise Spectrum at Home and a Lab.} % caption for whole figure
	\label{Fig:noise}
\end{figure}

The choice of frequencies in the two groups is also impacted by another factor. Although the new DTMF frequencies are a good fit for data exfiltration in our channel, we show in Section~\ref{subsubsec:error} that noise is the main source of errors when we transfer data with DTMF tones. The background noise spectrums in a normal office and home setting are shown in Figure~\ref{Fig:noise}. We can see that in the lab, the noise energy mostly distributes at lower frequencies below 600 Hz, while at home, the noise tends to distribute more randomly across the voiceband. 
%The frequency spectrums of these two environment are mainly affected by the HVAC systems. A more powerful air conditioner in the lab tends to emit more lower frequency noise. 
Also, while high frequencies (above $500$ Hz) tend to be very directional, lower frequencies can more easily spread and reflect in a room, resulting in increased low frequency noise in our recorded audios~\cite{sound:2008}. Based on these observations, the lowest frequencies chosen by us can avoid most of the interference coming from noise. Although the ranges of the two groups have considerable overlap, this does not influence our decoding for DTMF tones, since the frequencies in the two groups are separated by at least 70 Hz, and the two groups are detected separately by the Goertzel algorithm. The two new frequency groups also satisfy the criteria for avoiding harmonics lower than sixth as in~\cite{battista1963signaling}. Furthermore, it will avoid most collisions when we use high-frequency modulation to the tone audio before transmitting it over the computer speaker.

%we chose to add $\{1040, 1149, 1270, 1404\}$(Hz) to low group to expand the set of DTMF frequencies.

%For the high group, the frequency set can be expanded either by adding new frequencies below the $4\times4$ DTMF high group or above it. To minimize overlap with low group, we chose not to add below the DTMF existing high group. Therefore, we expand the high group by adding $\{1805, 1995, 2205, 2437\}$(Hz). 

The groups of new DTMF frequencies are shown in Table~\ref{Table:ext_code}. With expanded groups, we have $8\times8 = 64$ DTMF tones. Thus, each tone can now represent 6 binary bits, increasing the transfer bandwidth by 50\% compared with the commonly used DTMF tones. 

\begin{table}[!htb]
\centering
\caption{Expanded DTMF Frequencies.}
\label{Table:ext_code}
\resizebox{.3\textwidth}{!}{%
\begin{tabular}[t]{c|c}
Group A (Hz) & Group B (Hz) \\ \hhline{=|=}
1624            & 1402            \\
1794            & 1549            \\
1982            & 1712            \\
2190           & 1892            \\
2420            & 2091            \\
2674           & 2311            \\
2955           & 2554            \\
3266         & 2822            \\
\end{tabular}%
}
\end{table}

\subsection{Data to Audio Encoding/Decoding}
To transmit data, we must first convert the data that is to be transmitted to an audio file. The encoding method used should be such that the encoded audio can be transmitted via the computer speaker, received by a VA and then sent via a phone channel.

\subsubsection{Converting Binary Data to DTMF Tone Audio}

We now describe the process that starts from a data file that stores arbitrary data and outputs an audio file that can be played through the computer speaker. Binary values can be mapped to DTMF tone sequences in a straightforward manner. The DTMF generator generates one tone for every $\log_2(n)$ bits, where $n$ is the number of tones we are using. In widely used DTMF systems, $n=16$, so a tone can represent 4 bits. However, in our expanded DTMF, $n=64$, allowing one tone to represent 6 binary bits.  As discussed in Section~\ref{Sec:phone}, our characterization of the data transfer channel used by VAs demonstrates the need for handling transmission errors that could arise due to noise or other factors like packet losses. We develop an encoding scheme that allows us to handle certain class of transmission errors.  

\begin{figure}[htbp]
\centering
\includegraphics[width=2.7in]{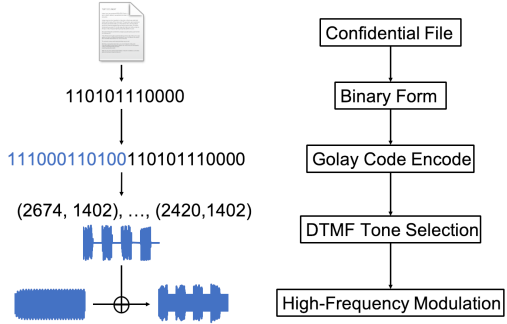}
\caption{Encoding Process. One 24-bit Golay code word is represented by 4 DTMF tones. }
\label{Fig:encode}
\end{figure}

As shown in Figure~\ref{Fig:encode}, we first convert the characters in a file we are sending to their corresponding ASCII binary codes. Thus, one character can be represented with 8 bits. For error correction, we first use the Golay code for encoding these bits before generating tones from them. The choice of this code is motivated by our desire to correct certain errors that can occur during transmission (see Section~\ref{subsubsec:error}).

\begin{table*}[]
\centering
\caption{Two Dimensional Gray Code Encoding DTMF}
\label{Table:gray}
\resizebox{.85\textwidth}{!}{%
\begin{tabular}{|c|c|c|c|c|c|c|c|c|c|}
\hline
              & Group A (Hz)   & 1624    & 1794    & 1982    & 2190    & 2420    & 2674    & 2955    & 3266    \\ \hline
Group B (Hz) & Gray Code      & 000     & 001     & 011     & 010     & 110     & 111     & 101     & 100     \\ \hline
1402          & 000            & 000 000 & 001 000 & 011 000 & 010 000 & 110 000 & 111 000 & 101 000 & 100 000 \\ \hline
1549          & 001            & 000 001 & 001 001 & 011 001 & 010 001 & 110 001 & 111 001 & 101 001 & 100 001 \\ \hline
1712          & 011            & 000 011 & 001 011 & 011 011 & 010 011 & 110 011 & 111 011 & 101 011 & 100 011 \\ \hline
1892          & 010            & 000 010 & 001 010 & 011 010 & 010 010 & 110 010 & 111 010 & 101 010 & 100 010 \\ \hline
2091          & 110            & 000 110 & 001 110 & 011 110 & 010 110 & 110 110 & 111 110 & 101 110 & 100 110 \\ \hline
2311          & 111            & 000 111 & 001 111 & 011 111 & 010 111 & 110 111 & 111 111 & 101 111 & 100 111 \\ \hline
2554          & 101            & 000 101 & 001 101 & 011 101 & 010 101 & 110 101 & 111 101 & 101 101 & 100 101 \\ \hline
2822          & 100            & 000 100 & 001 100 & 011 100 & 010 100 & 110 100 & 111 100 & 101 100 & 100 100 \\ \hline
\end{tabular}%
}
\end{table*}

\subsection{Error Correction}
\label{subsubsec:error}
%\subsubsection{Error Exploration}
We develop our error correction scheme by first exploring the source of errors in the received audio. Such errors occur because the audio of a tone is sufficiently perturbed such that a frequency that does not belong to the tone has higher energy level than the tone frequencies. In our experimental setting, there are two main sources of errors: Noise and the unwanted frequencies introduced by demodulation. We employ mitigation methods for each of them by using an error correction code, The Golay Code allows us to correct a certain set of errors.

\subsubsection{Errors from Noise}
\begin{figure}[htbp]
\centering
\includegraphics[width=3in]{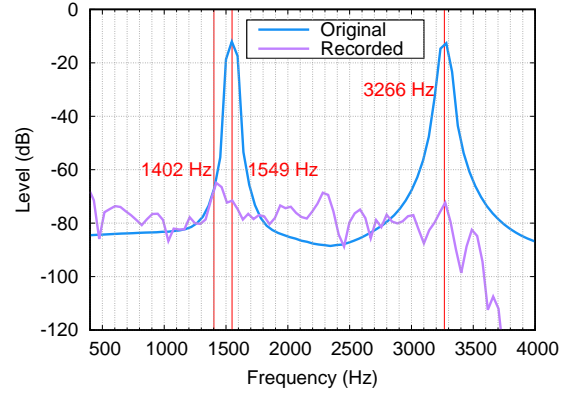}
\caption{Spectrum of Original Audio and Recorded Audio. We take a 0.05s tone from our experiment when the distance is 50 inch and carrier frequency is 18 kHz. Frequency 1549 Hz was decoded as 1402 Hz, due to frequency leakage and noise.}
\label{Fig:error}
\end{figure}

We empirically studied how such errors arise and found that the main reason is the background noise. In Figure~\ref{Fig:error}, the original tone consists of two frequencies, 1549 Hz and 3266 Hz. However in its recorded audio, a number of new frequencies appear due to background noise. In particular, the frequency 1402 Hz with higher energy level takes the place of original 1549 Hz, resulting in a bit flip in decoding. We did find that the rate of such errors can be reduced by using a lower carrier frequency or by using longer tones, which is a tradeoff between stealthiness and speed.

We also noticed that the impact of noise is such that errors that lead to bit flips are due to increased energy level of DTMF frequencies in adjacent tones (an adjacent tone encodes a bit string that differs in a single bit) caused by spectral leakage after Fourier transform. This can be seen from Figure~\ref{Fig:error} where we see that both in the transmitted and received audios, the energy of each DTMF frequency is distributed around this frequency. Thus, when the tone audio suffers additive noise, there is a higher probability that the adjacent frequencies energy exceeds the energy level of the tone frequency. Since decoding is based on energy levels of the tone frequencies, this will lead to a bit flip because an adjacent but incorrect frequency is detected. 
%To correct such errors, we use the two dimensional Gray code.

We use a two dimensional Gray code to encode data to DTMF frequencies. Gray code is a well-known binary code in which codes for two successive values differ only in one bit. %Gray code can be used to correct a single bit flip that may occur due to noise.
We can see in Table~\ref{Table:gray} that DTMF tones form a two dimensional matrix constructed from two frequency groups. Since each tone $T$ can have four different adjacent tones, each of which has one frequency different from $T$, we can apply the idea of Gray code to this matrix. First, with 8 frequencies, each frequency can be represented with a 3-bit Gray code. We can then combine two 3-bit codes together, with the Group A frequency bits at left and Group B frequency bits at right as in Table~\ref{Table:gray}. This results in a DTMF tone binary code matrix that satisfies the property  that adjacent tones differ only in one bit. We can write this table as a function with 6 binary bits as input and corresponding frequencies as output .
\begin{displaymath} T(B_i) = (F_A,F_B), \end{displaymath}
where $B_i$ is the $i$th 6-bit input binary string, $F_A$ is the Group A frequency of the DTMF tone and $F_B$ is the Group B frequency. When decoding the message, the function is the inverse of $T$:
\begin{displaymath}
S(F_A, F_B) = T^{-1}(F_A,F_B).
\end{displaymath}
For example, if we are sending binary string $B_1 = 001010$, corresponding DTMF tone $T_1 = (1794, 1892)$ will be sent through the phone call channel. If noise is added to the tone audio, adjacent tone frequencies could have higher energy levels. If the received tone is $T_1^{'}= (1624, 1712)$, the decoding function will produce the binary string $B_1^{'} = S(T_1^{'}) = 000011$. We can see that $B_1$ and $B_1^{'}$ have only 2 bits difference when noise impacts both tone frequencies.

\subsubsection{Errors from Demodulation}
We modulate the tone audio with carrier frequencies 15 kHz, 18 kHz or 20 kHz to achieve stealthy transfer that is less likely to be noticed by a person who may be in the vicinity of the computer. Due to the nonlinearity of voice assistant microphones, the modulated signal will get naturally demodulated before it goes to the phone channel. Because of the voiceband range (300--3400 Hz) of the phone call channel, the high frequency components that are close to the carrier frequency will be removed, which gives us the original signal. In Section~\ref{subsec:nonlinear}, we discover that the modulated sound will generate $f_1$, $f_2$ and an unwanted frequency, $(f_1-f_2)$. Since we choose $f_1 = f_A + f_{B}$ and $f_2 = f_A$, it produces $(f_A + f_{B})$, $f_A$ and $f_{B}$ in the recording. Thus, there are only 6 possible values of the additional frequency $(f_A + f_{B})$ that lie in voiceband and will  be present in the recording.
%\todo{Does the Goertzel algorithm not return these frequencies?}

%Although we avoid most collisions, the $(f_H + f_L)$ part still can introduce errors due to an imbalanced demodulation. We modulate all the DTMF tones and plot the frequency response of the recording inn Figure~\ref{Fig:demodu}. The figure shows that the recorded signal has obviously more energy level at higher frequency end. This trending will lead low frequencies being decoded to high frequencies. To minimize this kind of error, we compensate the loss of energy in low frequencies according to the frequency value in decoding algorithm. In stead of using the original energy level $G$, we empirically decided to compare the value 
%\begin{displaymath}
%G^{'}=G-50.3\log_{10}(f)+168.51,
%\end{displaymath}
%where $f$ is the frequency value. It gives more weight to lower frequencies. The red line in Figure~\ref{Fig:demodu} is the compensated frequency energy distribution. After applying this compensation, the number of error tones decreased from 8 to 3 out of 160 tones.

%\begin{figure}[htbp]
%\centering
%\includegraphics[width=2.7in]{pics/frequency_response.eps}
%\caption{Compensation of demodulated signal. The recorded sound has more energy distributed at higher frequency end. To compensate the low frequency in decoding, we compensate the gain according to the frequencies.}
%\label{Fig:demodu}
%\end{figure}

\subsubsection{Golay Code}

We use Golay Code for error correction. It can correct at most 3 bits errors in a 24-bit string~\cite{thompson1983error}. Golay code comes with an encoder $E()$ and a decoder $D()$. A 12-bit binary string message, $M$, will be encoded into a 24-bit Golay code word, $W = E(M)$. Since one DTMF tone can represent 6 bits of binary strings, one Golay code word is transmitted by 4 tones. 

After receiving the code, every 4 tones will be decoded into one Golay code word, $W^{'}$, which may have several flipped bits compared with the original word $W$. We can then get the message $M^{'} = D(W^{'})$. If the flipped bits are less or equal to three, $M^{'}=M$.

%\subsection{Sound Masking}

%We add additional cover sound to the encoded audio files to mask the tones in order to make them less noticeable. In our experiments, we use white noise and violin music to cover the generated audios. Violin music has a highest possible frequency, as shown in Table~\ref{Table:range}, and it is continuous. As we discussed in Section~\ref{sec:masking}, people will focus more on higher frequency and music, which is violin music in our experiments, without noticing the covered DTMF tones which encode the data.
%To ensure the masking audio does not overlap with the DTMF frequencies,  we use a high pass filter to filter out frequencies that are below $3,000$ Hz in the violin music, leaving the lower band for the DTMF tones. Since the data is transmitted over the voice band, the phone channel acts as a low pass filter and drops frequencies higher than $3,400$ Hz. Thus, the use of frequency masking audio can make DTMF tones less noticeable to humans without impacting decoding of the audio at the receiver end.

\subsection{Data Exfiltration}

We consider the scenario in which an attacker has successfully compromised a computer and wants to exfiltrate sensitive data that is available at the computer. To bypass network defenses, malware on the compromised computer uses a VA and the phone channel to transmit data to the attacker. To accomplish this, the malware must first locate the data and convert it to an audio file. This is done using DTMF tones and the encoding scheme described by us. Once this is done, the tone audio is modulated with a carrier frequency chosen to make its transfer via a computer stealthy. These steps have to be completed before transmission can begin by playing the audio file via the computer speakers.

\begin{figure}[htbp]
\centering
\includegraphics[width=2.7in]{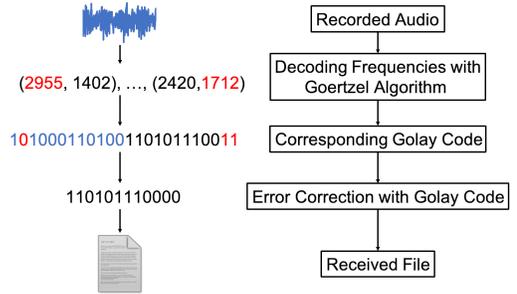}
\caption{Decoding Process. As shown in Figure~\ref{Fig:encode}, the first tone in the audio sent was (2674, 1402) and the last was (2420, 1402). They got decoded to (2955, 1402) and (2420, 1712), but the binary code is able to error correct this and decode the sent value correctly.}
\label{Fig:decode}
\end{figure}

To begin data transfer, first the VA must be given a command to make a phone call to a phone number chosen by the attacker. Such an activation command can be hidden and made unnoticeable as demonstrated by previous research~\cite{Yuan:2018um,Carlini:2016vl,Carlini:2018wj,Abdullah:2018ho,Roy:2018ws,Zhang:2017kl}. After the activation, desired audio will be played by the computer. The VA transmits this audio over the phone call and it is recorded at the receiving end of the call.
To recover the transferred data, prior to decoding, the audio in the recording is first aligned. This can be accomplished by adding a prefix to the audio. We then feed the audio to the Goertzel Algorithm to find out frequencies in each tone. Based on the frequencies found, a binary representation is generated. We use such representation as a sequence of 24-bit Golay code words, which are then decoded. Finally, the decoded binary data is  converted back to characters in the original file, as shown in Figure~\ref{Fig:decode}.

\section{Evaluation}
To evaluate the efficacy of  the data exfiltration scheme developed by us, we implemented a proof-of-concept prototype system. This system allowed us to demonstrate the feasibility of data exfiltration via a VA and the phone channel, and we  also used it  to experimentally assess how accurately data can be transmitted and what transfer rates can be achieved. In particular, we want to study the factors that impact error rates (e.g., incorrect characters in messages received in the phone recording at a phone number to which a call is made by the VA). The transfer accuracy is impacted by a number of parameters, some of which could be under the control of the attacker. Our experiments allowed us to study the impact of the following key parameters.

\begin{enumerate}
\item Tone length: Tone length describes the time duration for which a single tone is played. Shorter tones lead to faster data transfer but could also increase errors due to signal leakage across tones. 
%\item Audio volume: Higher volume makes it easy for the VA to capture and transmit the audio. However, increased volume also leads to higher noticeability in case a person is in the vicinity of the VA or computer.
%\item Masking sounds: Masking sounds reduce noticeability of audio tones but the sounds added to the audio may increase errors.
\item Distance between computer and VA: The computer from which data is exfiltrated could be several feets away from the VA. Increased distance allows greater flexibility for attacker but could increase transmission error rates because of reduced energy level of the audio signal that will be received at the VA.
\item Carrier frequencies: We modulate DTMF tone audio with a carrier frequency to achieve stealthiness. Higher carrier frequencies, which are closer to 20 kHz, are inaudible to almost all people and would not raise an alert. However, higher frequencies introduce more errors because they suffer higher attenuation with distance. We conduct experiments to  explore the stealthiness and transmission accuracy tradeoffs.
\item Background noise: We saw that additive background noise can increase the energy level of a frequency that is different than the tone frequencies. This leads to errors reducing accuracy.

\end{enumerate}

We report results of experiments that allowed us to explore the tradeoffs introduced by these parameters. We first describe the experimental setup and then present the results.

\subsection{Experimental Setup}
In our experiments, we use an Amazon Echo Plus to dial the attacker controlled phone number. Once the call is made, it can be recorded or allowed to go directly to voicemail. We give the command to make the phone call and play carrier modulated audio after it is set up from a Macbook Pro 15'' 2018 at 100\% volume. The audio goes over the air to Echo which is placed at various distances. It is then transmitted through the phone call channel and recorded using Google Voice. The attacker phone number is a Google Voice number, which allows voicemail or call recording to be downloaded from its website. When adding background sound, we play a night ambient noise from an iPhone 6 at 100\% volume at 5 inches away from the Amazon Echo Plus. 

We change the tone length from 50 ms to 12 ms to explore speed of transmission.
%(the standard en tone length is 40 ms with 20 ms gap). 
For measuring the possible attack range, the distance between the computer and the Echo varies from 0 inch to 100 inches. 
%We also change the volume at which the computer plays the audio to determine the lowest volume at which transmission is possible.
Also, different carrier frequencies between 15 kHz and 20 kHz to explore the tradeoff between stealthiness and accuracy. The influence of the presence of background sound is studied as well.

\begin{figure*}[!t]
	\centering
	\subfloat[The bit transmission accuracy of different tone lengths when Amazon Echo Plus is placed by the laptop and 25 inches away from it. The tones are modulated with 15 kHz, 18 kHz and 20 kHz carrier frequencies.\label{Fig:distance_length}]{\includegraphics[width=.3\linewidth]{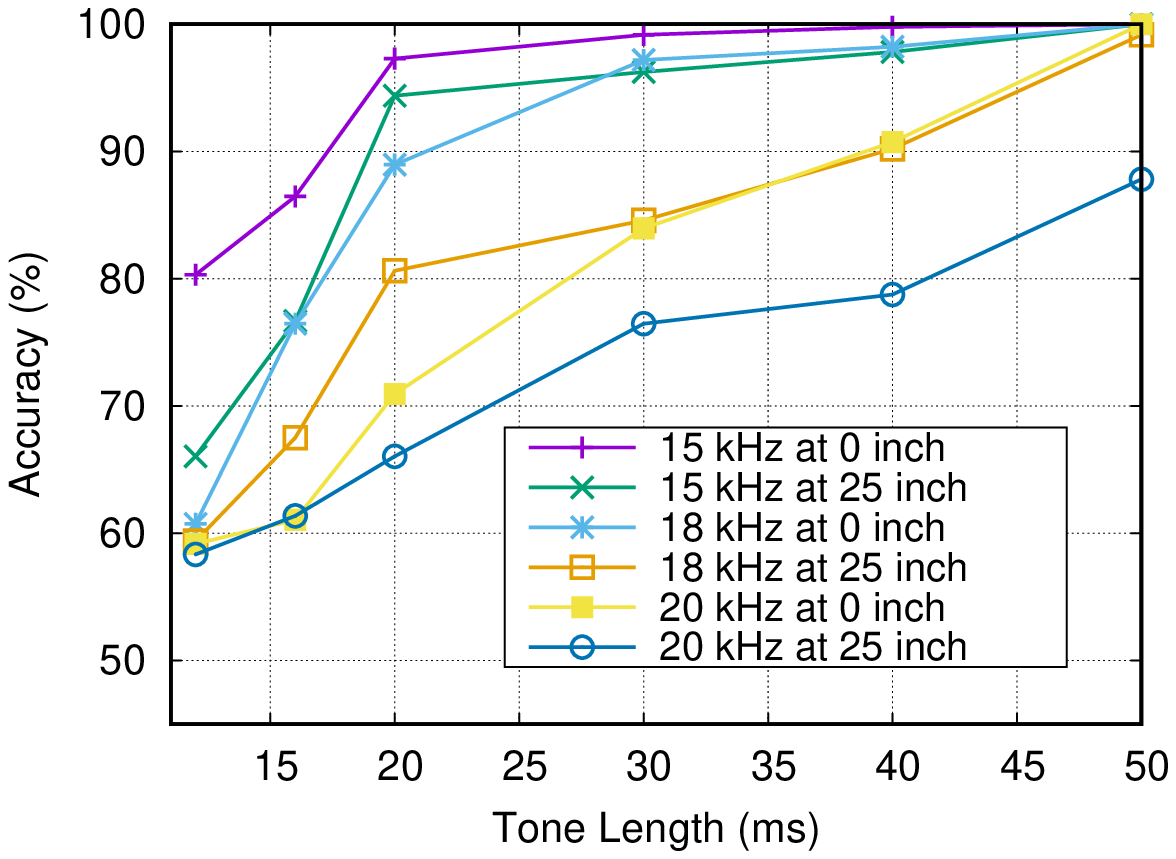}}
	\hspace*{\fill}  
	\subfloat[The number of errors after decoding for different distances when transmitting 40 characters. Tone length is 50 ms. Carrier frequency is 18 kHz.\label{Fig:error_correction_distance}]{\includegraphics[width=.3\linewidth]{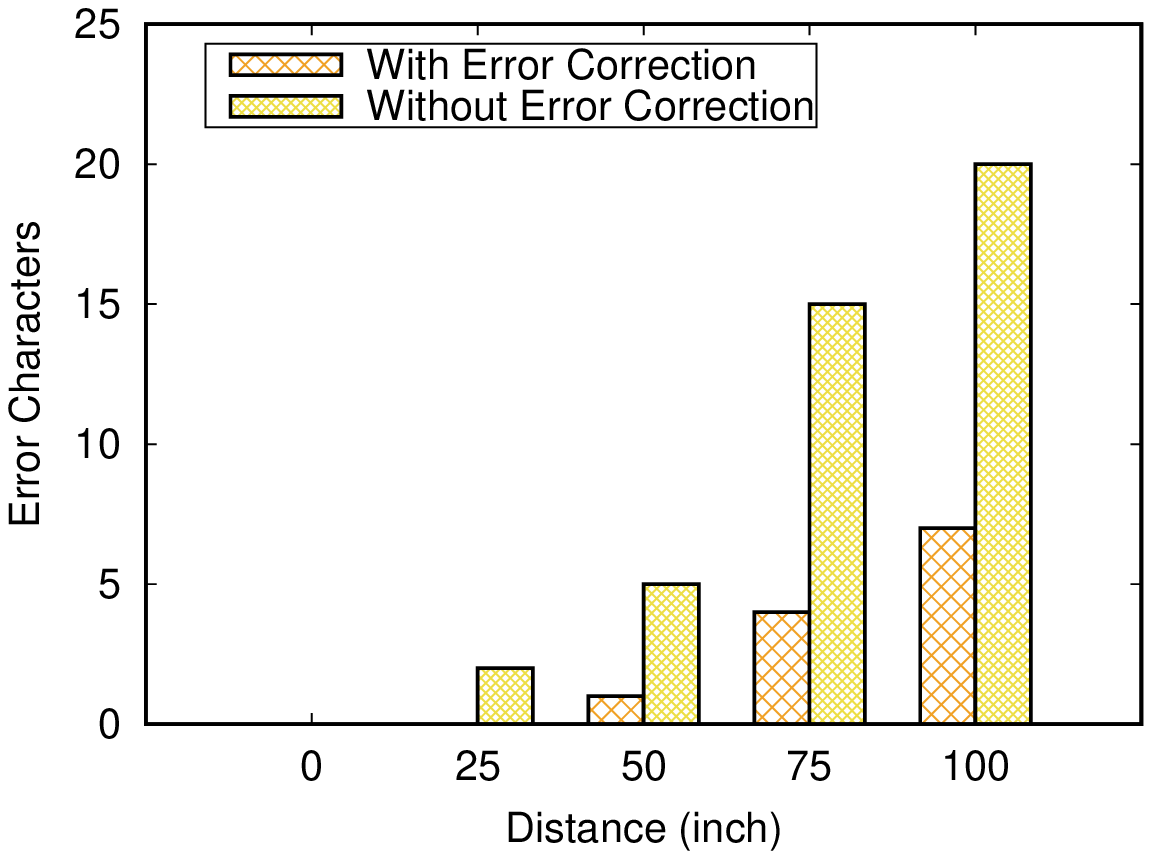}}
	\hspace*{\fill} % here you can insert horizontal or vertical space
	\subfloat[The number of errors after decoding for different tone lengths when transmitting 40 characters. Distance is 0 inch. Carrier frequency is 15 kHz. \label{Fig:error_correction_tone}]{\includegraphics[width=.3\linewidth]{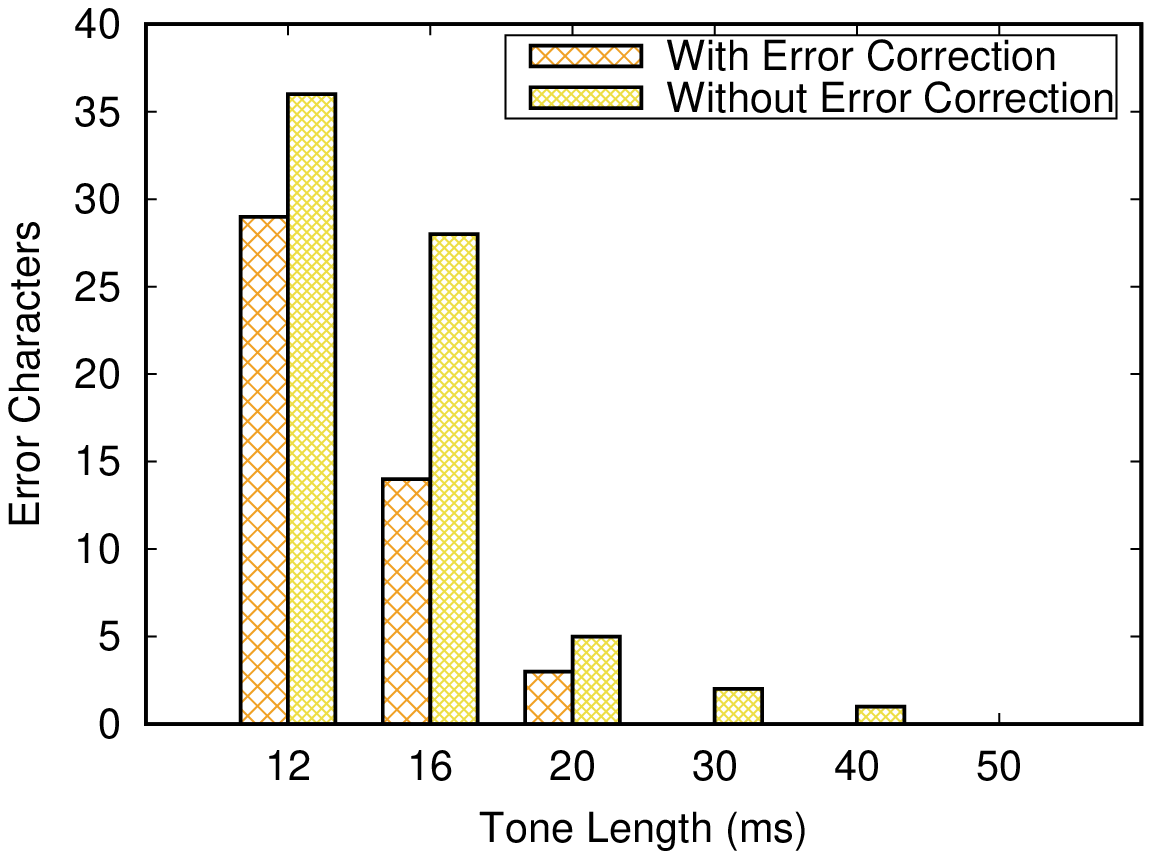}}

	\subfloat[The number of errors after decoding for different tone lengths at distance 0 inch when transmitting 40 characters. When tone is 50 ms long, there are no errors for all three carrier frequencies. \label{Fig:length_error}]{\includegraphics[width=.3\linewidth]{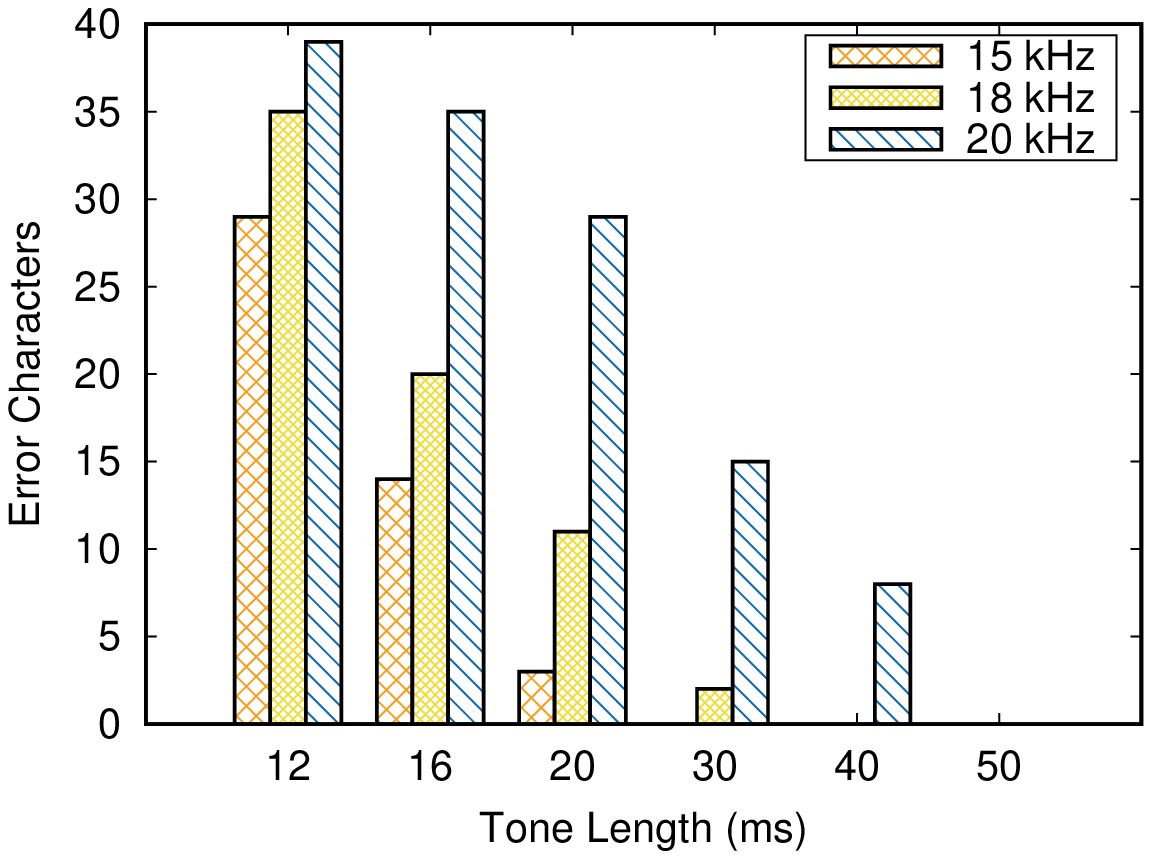}}
	\hspace*{\fill}  
	\subfloat[The bit transmission accuracy of different distances between laptop and Amazon Echo Plus. Tone length is 50 ms. Quiet means there was no background noise playing during transmission. Night Ambience means a night ambience sound audio was playing near Amazon Echo Plus. \label{Fig:volume_distance}]{\includegraphics[width=.3\linewidth]{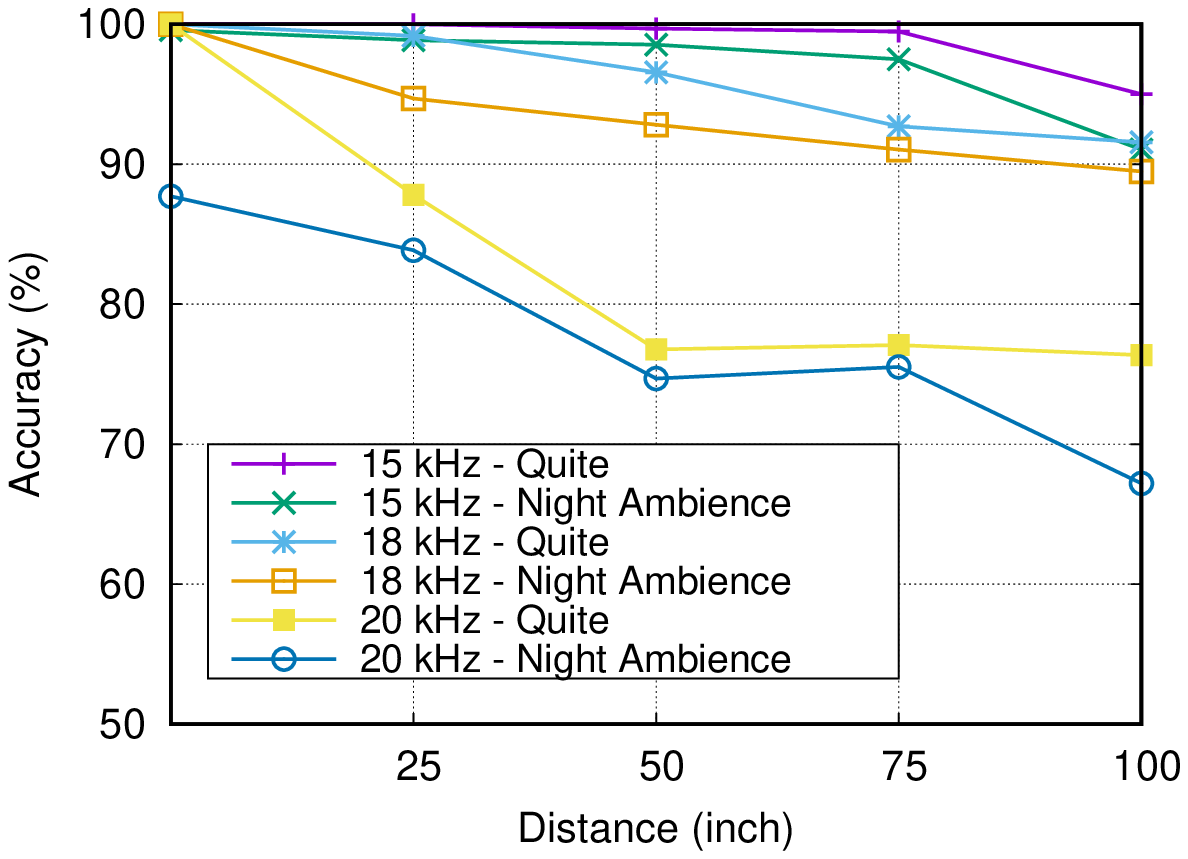}}
	\hspace*{\fill} % here you can insert horizontal or vertical space
	\subfloat[The number of errors after decoding for different distances when transmitting 40 characters. Tone length is 50 ms.\label{Fig:distance_error}]{\includegraphics[width=.3\linewidth]{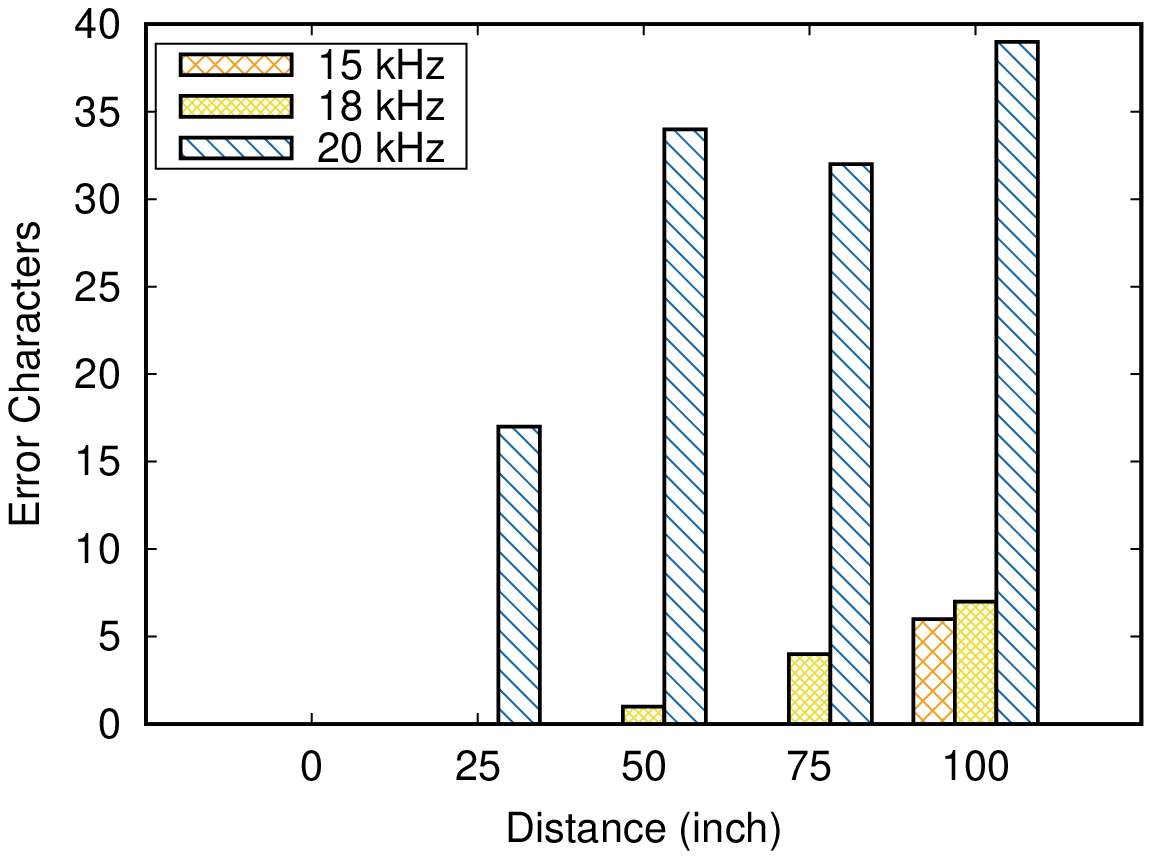}}
	\caption{Transmission Accuracy Results}
\end{figure*}

\subsection{Results}

We measure transmission accuracy under various settings of the parameters described above. If $n$ bits are transmitted and $m$ of those bits are received correctly, percent accuracy is defined as $m/n \times 100$.  Since we use 8 bits to encode characters, we also consider the number of characters that are received in error. We plot accuracy when different parameters are varied.
% and  masking sounds and noise are used to make the tone audio less noticeable.

\noindent
{\bf Tone Length:}
In Figure~\ref{Fig:distance_length}, we plot accuracy as it varies with tone length when various carrier frequencies are used to make the audio transmission less noticeable. As the tone length increases, the accuracy grows. This can be explained as follows. First, if the tone is too short, the adjacent tones will interfere with each other due to signal distortions introduced by the speaker and microphone; second, a shorter tone will result in lower amount of accumulated energy level for a frequency compared with noise, leading to incorrect decoding. When the tone length is longer than 40 ms, the accuracies are all above 90\% except for 20 kHz modulation at 25 inch distance. In this case, accuracy is lower as there are usually about $1/4$ character errors after error correction. When increasing the modulation frequency, we can see that the accuracy drops drastically when tone length is less than 20 ms. If we increase the tone length to 50 ms, we get closer to 90\% accuracy even with 20 kHz carrier frequency for modulation and  25 inches distance between the computer and the VA.  As can be seen, lines corresponding to carrier frequencies 15 kHz and 18 kHz go up to about 99\% with either distance. This demonstrates the tradeoff between stealthiness and transmission accuracy.

If the VA is very close to the computer, data can be transmitted with high accuracy. We can also see the number of characters that are received in error when the distance is 0 inch in Figure~\ref{Fig:length_error}. No characters are received in error when the tone length is 50 ms. From these experiments, we find that 50 ms tone length offers  good accuracy and we use this value in all other experiments. 

If the tone length is 50 ms and we leave a 50 ms gap between adjacent tones, there will be 10 tones/second. Each tone represents 6 binary bits, then the bit rate is 60 bits/second when no error correction is done. The highest bit rate is reached when the tone length is 12 ms, that is, a bit rate of 250 bits/second is achieved but many characters are received in error. Errors will be problematic when sensitive data such as passwords or social security numbers are exfiltrated. In case of normal text, it may be possible to correct the erroneous characters but we do not explore this here.

\noindent
{\bf Distance:} 
We varied the distance from 0 inch, where Echo is right in front of the laptop computer to 100 inches, where Echo is placed at one side of a bedroom while the laptop is at the other side. As expected, in Figure~\ref{Fig:volume_distance}, the accuracy worsens as we increase the distance between the Echo and the laptop. When modulating with 15 kHz and 18 kHz carrier, the accuracies are all above 90\% even at 100 inches. The errors are about 1/8 at 100 inches distance as can be seen in Figure~\ref{Fig:distance_error}. While 20 kHz modulation can reach 100\% accuracy at 0 inch without adding noise and nearly 90\% with noise playing, the accuracy goes down quickly as distance grows. Also, we can see that the night ambience noise is introducing errors in our received bits. 

\noindent
{\bf Modulation Frequency:} 
We discuss the impact of modulation frequency when the tone length is fixed to 50 ms. When the frequency is 15 kHz in Figure~\ref{Fig:volume_distance}, the accuracies at all distance are all close to 100\%, even when the distance is 100 inches. When we increase the frequency to 18 kHz, as shown in in Figure~\ref{Fig:volume_distance}, the accuracy is still above 89\% even at 100 inches distance. Only when the frequency reaches 20 kHz, the accuracy drops quickly after 0 inch. From these experiments, with an 18 kHz carrier, we can obtain good accuracy while the audio being nearly unnoticeable. If the distance is too far to obtain acceptable accuracy, the carrier only needs to be set to 15 kHz. With 15 kHz carrier, close to 100\% accuracy can be achieved, while it does leak some noticeable frequencies. In this case, we could only hear some faint sounds when the audio was played when the our ear was right next to the computer speaker. No noticeable sounds were heard a few feet away from the computer. This is consistent with the past research that shows that frequencies above 15 kHz can only be heard when the sound volume is high~\cite{Suzuki2003PreciseAF}.

\noindent
{\bf Noise:} As expected, the introduction of noise close to the VA has a negative impact on data transmission. As can be seen in Figure~\ref{Fig:volume_distance}, introduction of night ambient noise has significant impact on the transmission accuracy when 18 or 20 kHz carrier frequencies are used. In contrast, minimal impact is observed for 15 kHz carrier even when the distance between the computer and the VA is several feet.

\noindent
{\bf Error Correction: } 
We performed data decoding both with and without error correction. When the tone length is fixed to 50 ms, no noise is added, and distance is varied from from 0 inch to 100 inches, the results are shown in Figure~\ref{Fig:error_correction_distance}. When distance is 0 inch, there are no error in characters decoded from received audio. However, when distance is increased to 25 inches, there are errors due to bit flips. All of these errors can be handled by error correction. We also varied the tone length from 12 ms to 50 ms, with no added noise, fixing the distance at 0 inch. As shown in Figure~\ref{Fig:error_correction_tone}, when tone length is 12 ms, too many bit flips lead to errors both when correction is and is not performed. In contrast, tone length of 16 ms shows about 50\% fewer errors with error correction. There are no errors when tone length is 50 ms and accuracy is nearly 100\% in both cases. Since errors do occur at certain distances and tone lengths, we use error correction despite the fact that it halves the transfer rate.

\noindent
{\bf Summary:} Our experiments demonstrate the feasibility of data exfiltration via a VA even when it is several feet away from a compromised computer. This can be done even when humans are close to the computer or VA because most people are unable to hear frequencies higher than 15 kHz. Based on our results, we can see that when error correction is used, 9000 bits can be transmitted by a five minute phone call to a remote endpoint with very high accuracy. This can be done when a 18 kHz carrier is used to make the audio  completely unnoticeable and the distance between the computer and the VA is over six feet. As expected, noise and distance both reduce the accuracy of transmission.

\section{Discussion}
We have demonstrated that infected computers in the vicinity of VAs can bypass network defenses by using a VA mediated phone call to leak data. In this section, we briefly discuss possible defenses and their efficacy. We also discuss some limitations of the system developed by us. 

\subsection{Defenses}
Phone calls typically are made by users whereas data exfiltration by a VA requires the audio to originate from a computer speaker. For commands to VAs, liveness detection techniques have been developed. For example, it is demonstrated in~\cite{blue2018hello} that certain characteristic frequencies are present in audio that comes from a speaker which are not present when an audio source is a live person. These frequencies are low ($< 200$ Hz) and will not be transmitted by the telephony channel. Hence, defense based on such frequency detection has to be deployed at the VA. Furthermore, there may be legitimate cases when a call source is a computer. For example, a conferencing application running on a computer may command a VA to reach certain participants by making a phone call. Also, future AI based applications running on a computer may  use VAs in a conversation with remote parties that are reached via the phone (e.g., Google Duplex~\cite{google}). Again, a liveness detection defense cannot always be used.

Defenses against malware or compromised applications that abuse computer speakers can also be deployed on the computer where sensitive data is stored. These include access control for speakers and monitoring of audio stream going to the speakers which may indicate audio encoded data exfiltration. For example, an audio stream that consists of DTMF tones only, could raise an alarm. However, this could be evaded by using frequencies other than DTMF. Since a VA must send the audio stream to the cloud, another possible defense is to examine the audio stream for possible anomalies. A variety of audio stenography techniques have been investigated in the literature and their use may evade such defenses.

There are other defenses such as voice biometric verification such as those used by Google Home and Siri VAs. These defenses are primarily for commands that are issued to VAs. Although attacks such as voice conversion~\cite{ergunay2015vulnerability} and voice deepfakes~\cite{wallstreet:2019}
% add this link to the wall-street story about this. https://www.wsj.com/articles/fraudsters-use-ai-to-mimic-ceos-voice-in-unusual-cybercrime-case-11567157402
can be used, these will be more targeted attacks and would require access to voice samples of victims. In contrast, we have explored large scale attacks which could be more challenging when voice biometric verification is used..

\subsection{Limitations}
We demonstrated that an infected computer can command a VA to set up a phone call that can transmit sixty bits per second.  If we use error correction, thirty data bits can be sent in a second. Such a rate may not be sufficient to achieve massive data transfer or large scale data dumps. For example, a call lasting ten minutes can only exfiltrate approximately 2.5K bytes of data at this rate. Our experiments have not explored accuracy of data transfer over long duration calls.  Also, the audio may not be completely inaudible to all people when a carrier lower than 20 kHz is used. In particular, young children may be able to hear some sounds when they are very close to the computer or the VA. This may raise an alert. However, achieving completely inaudible audio transmission for everyone for large volumes of data over the voiceband of a telephony call is a problem that has not been fully addressed by this paper.

We use the phone call feature of Amazon Echo. Although this feature is useful and likely be provided by VAs, it is possible that in the future calls are only allowed to certain phone numbers (e.g., those in a contact list). In this case, an attacker would have gain access to the voice mail of an allowed phone number or find a way to get a phone number controlled by it to the contact list. This is not addressed by us.

\section{Conclusions and Future Work}
As voice assistants become common in the same physical areas as computers, they could create a new channel for data exfiltration. In particular, malware infected computers can bypass normal network and host defenses by using the voice assistant and the phone channel to send sensitive data to an attacker controlled computer. We demonstrated that modest amount of data can be exfiltrated by encoding the data in audio and transmitting it  via a voice assistant initiated phone call that lasts only a few minutes. By using techniques such as modulation with very high frequency carriers, it is possible to send the audio from the computer to the voice assistant in a way that it is unlikely to be noticed by a person who may be in the vicinity of these devices. These attacks are of concern because they can be mounted from anywhere, at scale and at low cost.

 In the future, we plan to explore if the transmission of data such as text can be made more efficient when it is sent over the voiceband of a telephony channel. This should  be done while making the transfer completely unnoticeable as has been demonstrated for commands issued to voice assistants. We briefly discussed a number of defenses that may be possible against data exfiltration via voice assistants but their efficacy is unclear. Also, detection and defenses against malware that may use the audio channel between the computer and voice assistants need to be investigated.

\bibliographystyle{IEEEtranS}
%\begin{thebibliography}{1}
%\end{thebibliography}

\bibliography{\jobname}

\end{document}